\documentstyle[prl,aps]{revtex}
\begin{document}

\draft
\twocolumn[
\widetext

\title{Phase transition in two-dimensional multicomponent
Bose liquid: a new type of quasi-long-range order}
\author{I.~V.~Tokatly}
\address{Moscow Institute of Electronic Engineering,
Zelenograd, 103498, Russia}

\date{\today}
\maketitle

\widetext
\begin{abstract}
\leftskip 2cm
\rightskip 2cm
The $\nu$-component two-dimensional (2D) Bose liquid is considered. It
is shown that the finite temperature phase transition exists in this
2D continuously degenerate multicomponent system. The new type of
quasi-long-range order occurs as a result of this transition. The
correlation function in ordered state is a product of the power-law
XY-universality class type function and the exponential function
corresponding to the $SU(\nu)$ invariant model. The transition
temperature is calculated for the diluted system. It is found that the
physical mechanism of the phase transition is changed when $\nu$
exceeds the critical value $\nu_c = 4$. The experimental realization
of the considered model is discussed briefly.
\end{abstract}
]
\narrowtext

According to Mermin-Wagner theorem \cite{Mermin} the true long-range
order does not exist in 2D continuously degenerate systems at $T \ne
0$. Nevertheless, it has been shown by Berezinskii \cite{Berez} and
independently by Kosterlitz and Thouless \cite{Kost} (BKT) that the
finite temperature phase transition exists in two-component $O(2) \sim
U(1)$ symmetrical systems. One-component Bose liquid and XY-model are
widely known examples of such systems, which are usually referred to
as XY-universality class systems. Phase transition in XY-universality
class systems consists of formation of the state with nonzero
transverse stiffness. The last one leads to the existence of the sound
collective excitation branch and power-law decay of the correlation
function below the critical temperature $T_{BKT}$
\begin{equation}
<\psi^*({\bf r})\psi({\bf r}')> \sim |{\bf r}-{\bf r}'|^{-\alpha}
\label{1}
\end{equation}
Formation of such power-law "long-range order" is accompanied by the
superfluidity in 2D one-component Bose liquid \cite{Kost}.

Concerning $N$-component continuously degenerate systems with $N
>2$ there is a well known statement that any phase transition is
absent and the correlation function decays exponentially in two dimensions
\cite{Barber,Patash}. Renormalization group arguments proposed by
Polyakov \cite{Polyakov} for $O(N)$ symmetrical model lead to the
following expression for the correlation length $R_c$:
$$
R_c \sim \exp{\left( \frac{2\pi J}{(N-2)T}\right) },
$$
where $J$ is "exchange" constant for effective $N$-component
Heisenberg model.

The purpose of the present letter is to demonstrate the possibility of
the finite temperature phase transition in some multicomponent 2D
continuosly degenerate systems. Namely, the $\nu$-component ($\nu \ge
2$) Bose liquid will be considered. For simplicity we fix our
attention on diluted systems, but general qualitative results will be
valid for arbitrary density.

We start from the usual functional integral representation for the
partition function $Z$:
\begin{equation}
Z = \int D{\bf \Psi}^*D{\bf \Psi}\exp{\left(-\int\limits_0^{1/T} d\tau
\int d^D r L\right)}.
\label{2}
\end{equation}
Lagrangian $L$ has the form:
\begin{equation}
L = -{\bf \Psi}^*\partial_{\tau}{\bf \Psi} + \frac{|\nabla {\bf
\Psi}|^2}{2M} + \frac{1}{2}t({\bf \Psi}^*{\bf \Psi})^2 -
\lambda {\bf \Psi}^*{\bf \Psi},
\label{3}
\end{equation}
where ${\bf \Psi}$ is $\nu$-component complex vector, $M$ is the boson
mass and $\lambda$ is a chemical potential. Then we assume that the
spatial range of the boson-boson interaction $r_0$ is much smaller than the
value $\sqrt{2M\lambda}$. This condition allows one to apply the usual
for diluted Bose systems procedure of interparticle interaction
renormalization \cite{Popov,Lozovik,Gor1}. For example,
one gets for 2D system:
\begin{equation}
t \approx \frac{4\pi}{M\ln(\varepsilon_0/\lambda)}; \qquad
\varepsilon_0 = \frac{1}{2Mr_0^2}
\label{4}
\end{equation}
Number of particles $n$ is determined by the expression:
$$
n = -T\frac{\partial}{\partial \lambda}\ln Z
$$
The symmetry group of the lagrangian (\ref{3}) is $U(\nu)$. Another
words, the form (\ref{3}) is invariant under the global transformations:
${\bf \Psi}' = U{\bf \Psi}$, $U^+U = 1$
($U$ is $\nu \times \nu$ unitary matrix). Let us represent the
field $\bf \Psi$ in the following form:
\begin{equation}
{\bf \Psi}(x) = \sqrt{n_0 + \pi(x)}e^{i\phi(x)}{\bf S}(x),
\qquad {\bf S}^*{\bf S}=1.
\label{5}
\end{equation}
Here $x\equiv\{\tau,{\bf r}\}$ and $n_0$ is "bare" condensate density,
firstly introduced by Popov for 2D one-component Bose gas
\cite{Popov}. The fields $\pi$ and $\phi$ describe the fluctuations of
module and phase of initial $\bf \Psi$-variables. The $\nu$-component
unit vector ${\bf S}$ corresponds to rotations in the intrinsic
isospin space. Any configuration ${\bf S}(x)$ can be obtained by the
local $SU(\nu)$ rotation of arbitrary unit constant vector ${\bf
\Psi}_0$. So vector ${\bf S}(x)$ can be written in the form:
\begin{equation}
{\bf S}(x) = e^{i\Lambda_a\theta^a(x)}{\bf \Psi}_0.
\label{6}
\end{equation}
Where $\Lambda_a$ ($a = 1,...,\nu^2-1$) are generators of $SU(\nu)$
transformations. The representation (\ref{5}) is useful at low
temperatures, where one has:
\begin{equation}
n_0 \approx \lambda/t \approx n,
\label{7}
\end{equation}
and $\pi \ll n_0$.

Substitution of (\ref{5}) into (\ref{3}) leads to the lagrangian:
\begin{eqnarray}\nonumber
L& = & -i\pi \partial_{\tau}\phi - (n_0+\pi){\bf S}^*\partial_{\tau}{\bf
S} + \frac{n_0}{2M}\{(\nabla \phi)^2 + |\nabla{\bf S}|^2  \\
& - &
2i\nabla\phi ({\bf S}^*\nabla{\bf S}) \} + \frac{(\nabla\pi)^2}{8Mn_0} +
\frac{1}{2}t\pi^2
\label{8}
\end{eqnarray}
The expression (\ref{7}) and the inequality $\pi \ll n_0$ have been used
in (\ref{8}).

Bose liquid with lagrangian (\ref{8}) is not superfluide at any
dimension D. Indeed, the full momentum ${\bf P}$, calculated in the
frame of reference moving with a velocity ${\bf v}_s$
$$
{\bf P}= -\frac{i}{M} <{\bf \Psi}^*\nabla{\bf \Psi}>_{{\bf v}_s}
$$
prove to be infinite due to the quadratic dispersion law of fluctuations
connected with a vector ${\bf S}$ rotation. So the dissipation caused by
the fluctuations of ${\bf S}$ destroys the superfluide flow and the
state with $\phi = \phi_0 + {\bf v}_s{\bf r}$ is unstable.
Nevertheless, the phase transition occurs at $T\ne 0$ in considered
system. This statement is evident for $D\ge 3$ where the true
condensate and long-range order are formed below the Bose condensation
temperature. To show the existence of the phase transition in the
case $D=2$ we consider the correlation function behavior at $|{\bf
r}-{\bf r}'| \to \infty$.

The correlation function of the ideal $\nu$-component Bose gas takes the
form:
\begin{equation}
G(R)\equiv <{\bf \Psi}^*(0){\bf \Psi}({\bf R})> \approx \nu\sum_{\bf k}
\frac{2MT}{k^2 + R_c^{-2}}e^{i{\bf kR}},
\label{9}
\end{equation}
where the correlation length $R_c$ is determined by the equation:
$$
n = \nu \sum_{\bf k}N_B\left(\frac{k^2+R_c^{-2}}{2M}\right)
$$
($N_B(E)=(e^{E/T}-1)^{-1}$ is Bose distribution function). One gets
for two-dimensional system:
\begin{equation}
G(R)= \nu \frac{MT}{2\pi}K_0\left( \frac{R}{R_c} \right) \sim
\sqrt{\frac{R_c}{R}} \exp{\left( -\frac{R}{R_c} \right)}
\label{10}
\end{equation}
and
\begin{equation}
R_c \sim \exp{\left( \frac{\pi n}{\nu MT} \right)}.
\label{11}
\end{equation}
The unordered (high temperature) phase of the interacted Bose gas with
the lagrangian (\ref{3}) has the similar behavior of the correlation
function. This situation is analogous to the one-component case
\cite{Popov}.

To calculate the function $G(R)$ at low temperatures we use the
representation (\ref{5}). The asymptotics of $G(R)$ at large distances
is determined by the static (classical) part of the lagrangian (\ref{8}).
Another words, only fluctuations of fields $\phi$ and ${\bf S}$ with
zero Matsubara frequency contributes to $G(R \to \infty)$.
Corresponding part of the full lagragian is
\begin{equation}
L =\frac{n_0}{2M}\{(\nabla \phi)^2 + |\nabla{\bf S}|^2 -
2i\nabla\phi ({\bf S}^*\nabla{\bf S}) \}.
\label{12}
\end{equation}
Shift of the $\phi$-field
\begin{equation}
\phi = \phi + \psi, \qquad \psi = i\sum_{{\bf q,k}} \frac{{\bf
qk}}{q^2}{\bf S^*_{{\bf k+q}}}{\bf S_{{\bf k}}}e^{i{\bf qr}}
\label{13}
\end{equation}
leads to the result:
\begin{equation}
L =\frac{n_0}{2M}\{(\nabla \phi)^2 + |\nabla{\bf S}|^2 -
|{\bf S}^*\nabla{\bf S}|^2 \}.
\label{14}
\end{equation}
It should be mentioned that $\nabla\psi({\bf r})$ (\ref{13}) is the
potential part of the vector $i{\bf S}^*\nabla{\bf S}$. The solenoidal part
of this vector has not coupled with $\phi$-field due to integration
over spatial variables in (\ref{2}). Besides, $\psi({\bf r})$ is a
real function due to the condition ${\bf S^*S}=1$.

According to (\ref{14}) phase $\phi$ and isospin $\bf S$ fluctuate
independently. So the partition function can be written in the form:
\begin{equation}
Z = Z_{\phi}Z_{\bf S}.
\label{15}
\end{equation}
The first factor $Z_{\phi}$ corresponds to the partition function for $U(1)$
invariant system, that is equivalent to the two-dimensional Coulomb gas
\cite{Kost,Popov} with power-law behavior of the low temperature
correlation function. Thus, the correlation function of initial
$\nu$-component Bose liquid $G(R)$ takes the form (at low temperature):
\begin{eqnarray} \label{16}
G(R) &\sim & R^{-\alpha}<{\bf S}^*(0){\bf S}({\bf R})> \sim R^{-\alpha}
\exp\left(-\frac{R}{R'_c} \right) \\  \nonumber
\alpha & = & \frac{MT}{2\pi n_0},\qquad R'_c \sim \exp{(\frac{\pi
n_0}{(\nu-1) MT})}.
\end{eqnarray}
The approach developed by Polyakov \cite{Polyakov} (see also
\cite{Patash}) was used to estimate the correlator $<{\bf S}^*(0){\bf
S}({\bf R})>$ for our $SU(\nu)$ symetric model with partition function
$Z_{\bf S}$ (\ref{15}). Correlation function (\ref{16}) corresponds to
"ordered" state with nonzero transverse stiffness $n_0$. This function
is intermediate between the power-law Kosterlitz-Thouless correlation
function (\ref{1}) and an exponential function. Comparison of the high
temperature (\ref{10}), (\ref{11}) and low temperature (\ref{16})
expressions for $G(R)$ shows that the phase transition with change of
the correlation function should exist at some nonzero temperature.

There are two possible mechanisms of this transition. The first one is
Kosterlitz-Thouless mechanism connected with the vortices unpairing
process. Indeed, the system with the partition function $Z_{\phi}$ is
isomorphic to the 2D charged gas with logarithmic interaction, which
demonstrates the phase transition of that type at the temperature
$T_{BKT}$. The stiffness parameter at the transition point $n_0 (T_{BKT})$
and the critical temperature $T_{BKT}$ are related by the universal
Nelson-Kosterlitz equation \cite{Nelson}:
\begin{equation}
\frac{n_0(T_{BKT})}{2MT_{BKT}} = \frac{1}{\pi}.
\label{17}
\end{equation}
It should be mentioned that this transition appears within the
approximation (\ref{12}-\ref{14}), when the fluctuations of
$\pi$-field where considered to be weak (the condition $\pi \ll n_0$
has been used). According to the terminology used in \cite{Patash},
we call such approximation as "harmonic". The unharmonic vortexless
fluctuations can drive the phase transition either. That was
demonstrated by variational calculations for XY-model in
\cite{Patash}. The one-component Bose-liquid variational theory
considering selfconsistently the exact relationship between anomalous
$\Delta$ and normal $\Sigma$ self-energy functions ($\Sigma-\Delta =
\lambda$) has been developed in \cite{Gor1}. This theory allows one to
describe correctly the vortexless fluctuations in a dilute limit and
leads to the phase transition at some temperature $T_0$. However, the
unharmonic transition critical temperature $T_0$ for both XY-model
\cite{Patash} and one-component Bose liquid \cite{Gor1} prove to be
more than the BKT critical temperature ($T_0 > T_{BKT}$). It will be shown
bellow that the multicomponent Bose liquid represents the example of
the system, which demonstrates the phase transition of just unharmonic
nature.

The simplest extension of the approach \cite{Gor1} on the $\nu$-component
system leads to the following selfconsintecy equation for the anomalous
self-energy $\Delta$:
\begin{equation}
\Delta = tn - \nu\sum_{\bf k}\frac{k^2}{2ME_{\bf k}}N_B(E_{\bf k}),
\label{18}
\end{equation}
where
$$E_{\bf k}=\sqrt{\frac{k^2}{2M}\left(\frac{k^2}{2M} + 2\Delta
\right)}
$$
is the Bogoliubov dispersion law and $t$ is the renormalized interaction
(\ref{4}). The condition $\Delta/T \ll 1$ takes place for the diluted
system near the critical point. The stiffness parameter $n_0$ is
connected with $\Delta$ by the relation $tn_0 \approx \Delta$
\cite{Gor1}. Thus the equation determined $n_0$ near the
transition point takes the form:
\begin{equation}
n_0 = n - \nu \frac{MT}{2\pi}\ln\left(\frac{T}{2tn_0}\right).
\label{19}
\end{equation}
The solution of the equation (\ref{19}) appears abruptly with lowering
of $T$ at the temperature $T_0$:
\begin{equation}
T_0 =\left. \frac{2\pi n}{\nu M}\right/ln\left\{ \frac{e}{4\nu}
ln\frac{1}{r_0^2 n} \right\}.
\label{20}
\end{equation}
The relation between the stiffness and the critical temperature at the
unharmonic transition point looks similar to one at the
Kostrlitz-Thouless critical point (\ref{17}), but with an addition factor
connected with the number of components $\nu$:
\begin{equation}
\frac{n_0(T_0)}{2MT_0} = \frac{\nu}{4\pi}.
\label{21}
\end{equation}
Comparison of the formulas (\ref{21}) and (\ref{17}) shows that the
temperature of the Kosterlitz-Thouless phase transition $T_{BKT}$ is
less than the unharmonic transition temperature $T_0$ if the number of
components $\nu < 4$. The opposite inequality $T_{BKT}>T_0$ takes
place when $\nu > 4$. The value $\nu = 4\equiv \nu_c$ corresponds to
the critical number of components at which the critical points for
both of transitions are coincided and physical mechanism of the phase
transition is changed. To estimate the temperature $T_{BKT}$ one can
use the equation (\ref{19}) and the relation (\ref{17}). The result is
\begin{equation}
\frac{T_0-T_{BKT}}{T_0} =\left.\ln \left( \gamma e^{\gamma -1} \right)
\right/\ln\left\{ \frac{e^{\gamma}}{16} ln\frac{1}{r_0^2 n} \right\},
\label{22}
\end{equation}
where $\gamma = \nu_c/\nu$. The last formula directly demonstrates the
above mentioned behavior. It should be pointed out that the width of
the temperature region between $T_{BKT}$ and $T_0$ decreases with
density lowering. So it is possible to estimate the transition
temperature by the formula (\ref{20}) in dilute limit at any $\nu$.

In conclusion the existence of the phase transition in multicomponent
two-dimensional Bose liquid has been demonstrated. The correlation
function of the low temperature phase is the product of the power-low
Kosterlitz-Thoules type function and the exponential function corresponding
to $SU(\nu)$ invariant model. So the quasi-long-range order is formed
below the critical temperature. Let us discuss briefly the possible
experimental realization of the considered model. It may be a two
dimensional exciton liquid in semiconductor nanostructures with holes
from the $\Gamma$-point and electrons from the X-point of the Brillouin
zone. Manifestations of Bose-condensation in such system have been
reported recently \cite{Butov}. Most likely that the transition
observed \cite{Butov} is of the type described above.

This work has been supported in part by Russian Basic Research
Foundation and Russian Program "Physics of Solid State
Nanostructures".


\begin{thebibliography}{99}

\bibitem{Mermin} N.~D.~Mermin, H.~Wagner, Phys. Rev. Lett. {\bf 22},
1133 (1966).

\bibitem{Berez} V.~L.~Berezinskii, Zh. Eksp. Teor. Fiz. {\bf 61}, 1144
(1971) [Sov. Phys. JETP {\bf 34}, 2147 (1972)].

\bibitem{Kost}  J.~M.~Kosterlitz, D.~Thouless, J. Phys. {\bf C 6},
1181 (1973); J.~M.~Kosterlitz, J. Phys. {\bf C 7}, 1046 (1974).

\bibitem{Barber} Michael~N.~Barber, Physics Reports {\bf 59}, N~4
(1980).

\bibitem{Patash} A.~Z.~Patashinskii, V.~L.~Pokrovskii, {\it
Fluctuation Theory of Phase Transitions}, 2nd ed., Nauka, Moscow
(1982) [Pergamon Press, Oxford, 1979].

\bibitem{Polyakov} A.~M.~Polyakov, Phys. Lett. {\bf B 59}, 79 (1975).

\bibitem{Popov} V.~N.~Popov, {\it Functional Integrals in Quantum
Field Theory and Statistical Physics}, Atomizdat, Moscow (1976)
[Reidel, Dordrecht, 1983].

\bibitem{Lozovik} Yu.~E.~Lozovik, V.~I.~Yudson, Physica {\bf A 93},
493 (1978).

\bibitem{Gor1}  A.~A.~Gorbatsevich, I.~V.~Tokatly, Zh. Eksp. Teor.
Fiz. {\bf 108}, 1723 (1995) [Sov. Phys. JETP {\bf 81}, 945 (1995)];
Zh. Eksp. Teor.
Fiz. {\bf 108}, 2084 (1995) [Sov. Phys. JETP {\bf 81}, 1136 (1995)].

\bibitem{Nelson} D.~R.~Nelson, J.~M.~Kosterlitz, Phys. Rev. Lett.
{\bf 39}, 1201 (1977).

\bibitem{Butov} L.~V.~Butov, A.~Zrenner, G.~Abstreiter {\it et al}
Phys. Rev. Lett. {\bf 73}, 304 (1994).

\end{thebibliography}
\end{document}